%% file: main.tex
\documentclass[final,5p,times,twocolumn]{elsarticle}
\usepackage{graphicx} 
\usepackage{amsmath}
\usepackage{array}
\usepackage{multirow}
\usepackage{tabularx}
\usepackage{siunitx}
\usepackage{caption}
\usepackage{subcaption}
\usepackage{hyperref}
\hypersetup{
    colorlinks=true,
    linkcolor=blue,
    filecolor=magenta,      
    urlcolor=cyan,
    pdftitle={Overleaf Example},
    pdfpagemode=FullScreen,
    }
\usepackage{amsmath} 
\usepackage{color}
\definecolor{red}{rgb}{1.0,0.,0.0}
\usepackage[acronym,nomain]{glossaries} 
\setacronymstyle{long-short-desc}

\makeglossaries
\loadglsentries{glossary.tex}

\begin{document}

\begin{frontmatter}

\title{Hydrogen Inventory Simulations for PFCs (HISP)}

\author[inst1]{Kaelyn Dunnell\corref{cor1}}
\author[inst2]{Adria Lleal}
\author[inst3]{Etienne Augustin Hodille}
\author[inst3]{Jonathan Dufour}
\author[inst1]{Remi Delaporte-Mathurin}
\author[inst2]{Tom Wauters}

\cortext[cor1]{kaelyn@mit.edu}
\address[inst1]{Plasma Science and Fusion Center, MIT, Cambridge, USA}
\address[inst2]{ITER Organization, Route de Vinon-sur-Verdon, CS 90 046, 13067 St Paul Lez Durance Cedex, France}
\address[inst3]{CEA, IRFM, F-13108 Saint-Paul-lez-Durance, France}

\begin{abstract}
Hydrogen Inventory Simulations for Plasma facing components (HISP) is an open-source simulation tool to model the evolution of hydrogen (H) isotopes inventory in plasma-facing-components (PFCs) of magnetic confinement fusion devices. 
The objective was to produce a demonstrative study describing the efficiency of tritium (T) removal strategies in ITER. 
HISP transforms plasma code outputs to spatial-averaged inputs along ITER's first wall (FW) and divertor for 1D H transport models using FESTIM. 
Exposure conditions were tested in three scenarios that included DT operation and varied T removal methods. 
Generally, DT operation resulted in $\approx$ \SI{35}{g} of T in FW and divertor components after 10 days of DT pulses.
Almost \SI{80}{\%} of the total T inventory resided in co-deposited boron layers in the divertor. 
Baking proved to be the most effective T removal method in the divertor, decreasing T inventory by almost \SI{88}{\%} for tungsten and almost \SI{30}{\%} for boron.
T removal was also evaluated from Glow Discharge Conditioning (GDC)  - with a peak efficiency of \SI{23}{\%} in the tungsten FW - and low power deuterium (DD) pulses - with a peak efficiency of \SI{13}{\%} in the entire divertor. 
Due to the high removal efficiency of baking, inclusion of GDC and DD pulses in the tested scenarios did not meaningfully change final T inventory values, which varied by less than \SI{2}{\%} in the FW and \SI{10}{\%} in the divertor between scenarios. 

\end{abstract}


\end{frontmatter}

\section{Introduction}
Managing tritium (T) transport in fusion reactors is critical not just for fuel recovery and recycling efforts, but also for radiological safety. Due to tritium's radioactive nature, fusion reactors must anticipate tight tritium regulations \cite{nrc-regulate} aimed at securing the safety of fusion devices. The International Thermonuclear Experimental Reactor (ITER), for instance, has an in-vessel tritium inventory limit of \SI{1}{kg} at any given time; accounting for \SI{120}{g} of tritium trapped in divertor cyropumps and \SI{180}{g} of measurement uncertainty, this number drops to an actual limit of \SI{700}{g} for ITER \glspl{pfc} \cite{iter-t-limit}. Safety concerns and fusion efficiency require study of tritium removal strategies, through both computational and experimental means. 

\begin{figure}[h!]
    \centering
    \includegraphics[width=0.5\textwidth]{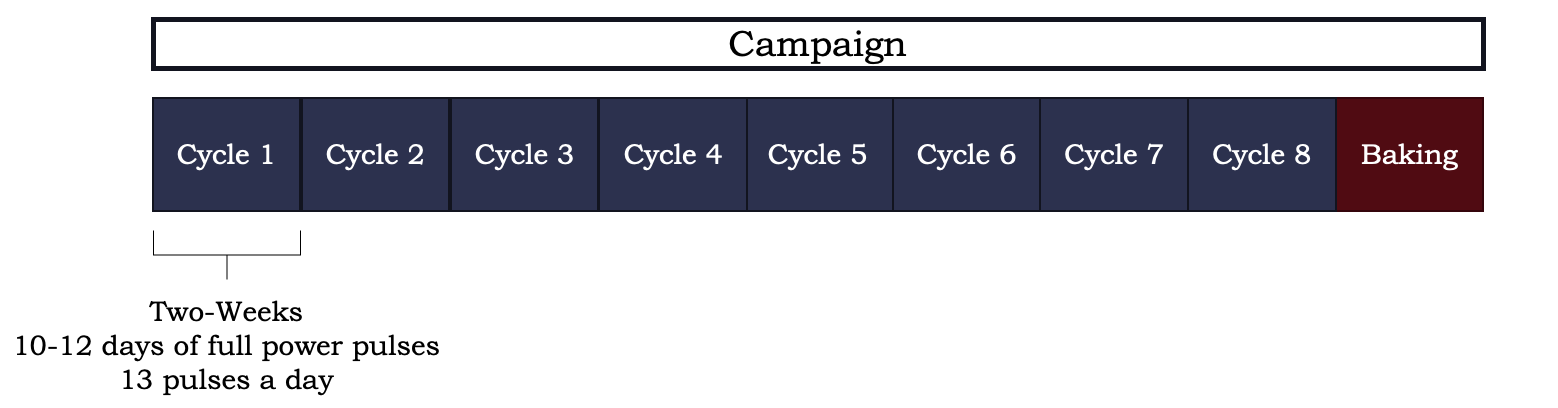}
    \caption{Illustration of an ITER Fusion Power Operation Campaign. A 16 month ITER campaign is broken into two-week cycles which contain 10-12 days of plasma operation and 2-4 days of short term maintenance. T removal techniques can be applied during either phase. The example campaign is concluded by a long period of baking applied during long term maintenance.}
    \label{fig:cycle-timeline}
\end{figure}

In ITER, several T removal techniques are envisaged to recover trapped T from \glspl{pfc} including baking and low temperature plasma methods~\cite{Loarte_2007}.
Such techniques can be applied in several instances during an operation campaign. ITER campaigns composed of two-week cycles (see Figure~\ref{fig:cycle-timeline}) of 10-12 days of plasma operation, with an average rate of 13 successful pulses per day, followed by \gls{stm} periods during which the toroidal magnetic field can be stopped \cite{iter-baseline}. The campaign concludes with a \gls{ltm} phase during which several weeks may be dedicated to T removal.  

Existing research has identified several tritium removal techniques \cite{wauters-wall-conditioning}. 
Pure deuterium (D) plasmas can be used to trigger thermal release or isotope exchange in the areas where T is retained~\cite{jet-post-experimental-campaign, Loarer_2013, hodille_2025}. To access specific areas, the plasma magnetic equilibrium can be adjusted relative to the ITER baseline scenarios; for example, shifting the position of the strike lines to target different surface regions.~\cite{jet-post-experimental-campaign}.
\gls{icwc} can be applied during operation days for tritium removal~\cite{WAUTERS2015}. \gls{gdc} requires the toroidal field coils to be de-energized, which is possible during \gls{stm}. Baking is a lengthy procedure which thus far is scheduled only during \gls{ltm}.  The aforementioned techniques have been demonstrated in the Joint-European Torus (JET) \cite{WAUTERS2015, jet-post-2nd-campaign, jet-post-experimental-campaign} and Tore-Supra/WEST~\cite{Loarer_2013,hodille_2025}. Although found effective, it remains unclear how the JET and WEST results extrapolate to ITER due to differing geometries. Therefore, developing a computational model to predict tritium inventory growth and evaluate removal strategies is essential to guide ITER operations. The goal of this research is to numerically find the most promising cleaning scenario for ITER with current knowledge. To accomplish this goal, Hydrogen Inventory Simulations for PFCs (HISP) was developed. 

\begin{figure*}
    \centering
    \includegraphics[width=\linewidth]{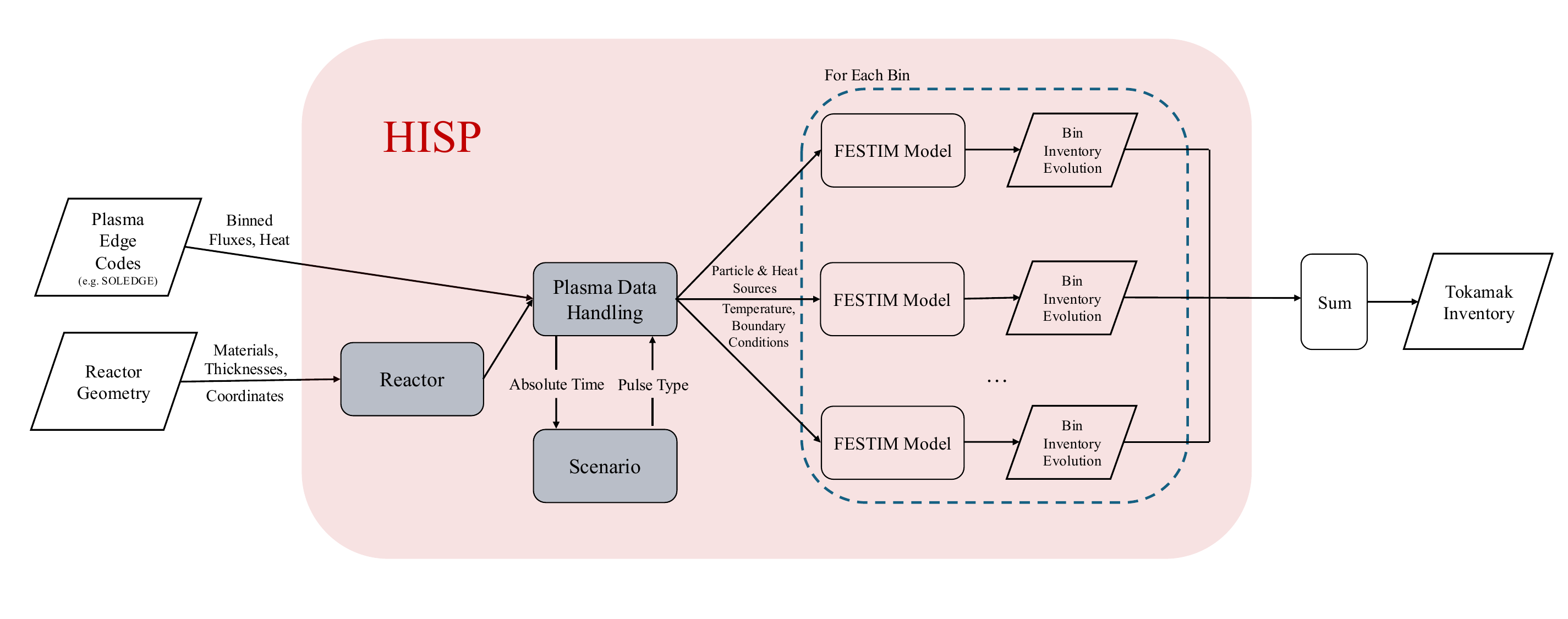}
    \caption{HISP workflow. See Figure \ref{fig:iter-bins} for bins.}
    \label{fig:hisp}
\end{figure*}

HISP is open-source, bridging the gap between FESTIM v2.0 \cite{dark_festim_2026} and plasma edge codes like SOLEDGE3X-EIRENE~\cite{Bufferand_2021} and SOLPS-ITER~\cite{WIESEN2015480}, which provide  particle flux and heat flux values associated with \gls{fw} and divertor components (see Figure \ref{fig:hisp}). Hydrogen transport codes like FESTIM, on the other hand, require heat and particle sources to estimate inventory evolution. HISP is the necessary middle step, transforming the outputs of plasma codes into spatial-averaged values (using ``bins'') for analysis by FESTIM. A HISP simulation requires description in three ways. First, the scenario to test; second, the plasma data to organize; and third, the location and material of the area of interest. Using this information, HISP produces hydrogen inventory studies for PFCs. In this paper, the hydrogen transport physics used in FESTIM will be described, as well as HISP's implementation strategy in detail and an example application to ITER. 

\section{HISP Methodology}
\subsection{Hydrogen transport with FESTIM}
\label{sec:HTMFES}
HISP leverages existing hydrogen transport simulation tools. In particular, this work proposes to use FESTIM \cite{dark_festim_2026} to simulate the transport of hydrogen isotopes in PFCs. FESTIM has a wide capability to describe hydrogen transport in materials as based on bulk and surface physics. Specifically, HISP is built on FESTIM v2.0, which includes features such as multi-isotope transport.  This paper will describe the physical processes used in HISP; for a complete discussion of FESTIM v2.0's simulation capabilities, see \cite{dark_festim_2026}.

FESTIM is based on the McNabb and Foster model \cite{mcnabb_new_1963} where the mobile and trapped hydrogen concentrations are described separately. The mobile deuterium concentration, $c_{\mathrm{D}}$ [\si{m^{-3}}], is described as 

\begin{equation}
    \label{eq:mobile-hydrogen}
    \frac{\partial c_{\mathrm{D}}}{\partial t} = \nabla \cdot (D \nabla c_{\mathrm{D}}) + S_\mathrm{D} - \sum_i{R_i}
\end{equation}

with the same equation following for the mobile tritium concentration, $c_{\mathrm{T}}$. The concentration of D trapped in trap $i$, $c_{\mathrm{D,t},i}$ [\si{m^{-3}}], is described as 

\begin{equation}
    \label{eq:trapped-hydrogen}
    \begin{split}
         \frac{\partial c_{\mathrm{D,t},i}}{\partial t} & = k_i c_{\mathrm{D}} (n_i - c_{\mathrm{D,t},i} - c_{\mathrm{T,t},i}) - p_i c_{\mathrm{D,t},i} 
         \\
         & = k_ic_{\mathrm{D}}n_{i,empty}-p_i c_{\mathrm{D,t},i}
         \\
         & = R_i
    \end{split}
\end{equation}

The same is true for the trapped tritium concentration. In these equations, hydrogen flux $J = -D \nabla c_{\mathrm{D}}$ with $D$ [\si{m^2.s^{-1}}] as the diffusivity coefficient represents Fick's Law of diffusion; $S_\mathrm{D}$ [\si{m^{-3}.s^{-1}}] is the source of mobile D; trap $i$ has trapping rate $k_i$ [\si{m^3.s^{-1}}], detrapping rate $p_i$ [\si{s^{-1}}], and trap density $n_i$ [\si{m^{-3}}]. The diffusion coefficient, trapping rates, and detrapping rates are described by the Arrhenius Law: 

\begin{equation}
    \label{eq:diffusion-coeff}
    D = D_{0} \exp \bigg[-\frac{E_{\mathrm{D}}}{k_\mathrm{B}T} \bigg]
\end{equation}

\begin{equation}
    \label{eq:arrhenius-trapping}
    k_i(T) = k_{0,i} \exp \bigg[-\frac{E_{\mathrm{t},i}}{k_\mathrm{B}T} \bigg]
\end{equation}

\begin{equation}
    \label{eq:arrhenius-detrapping}
    p_i(T) = p_{0,i} \exp \bigg[-\frac{E_{\mathrm{dt},i}}{k_\mathrm{B}T} \bigg]
\end{equation}

where $k_{0,i}$ [\si{m^3.s^{-1}}] and $p_{0,i}$ [\si{s^{-1}}] are the pre-exponential factors of the trapping and detrapping processes for trap $i$ (material dependent), $E_{\mathrm{t},i}$ is the trapping energy in eV, $E_{\mathrm{dt},i}$ is the detrapping energy in eV, and $k_B$ is the Boltzmann constant in \si{eV.K^{-1}}. 

Different isotope exchange mechanisms exist depending on the process \cite{hodille_2025}: (i) thermal detrapping, (ii) isotope swapping, or (iii) ballistic detrapping.
Isotope exchange through thermal detrapping is represented in the current model via Equation~\ref{eq:trapped-hydrogen}, where a D/T atom can exchange with a T/D atom at a trapping site only after the detrapping of the first atom.
Isotope swapping occurs when a D/T atom swaps directly with a T/D atom without the detrapping of the first atom ~\cite{Kogut_2016,Sun_2024}, a process inspired by atomistic calculations that suggest a reduction of the detrapping energy upon cumulative trapping~\cite{FERNANDEZ2015}.
HISP does not take this process into account either because of the lack of experimental evidence for this process (in boron especially), or because the traps in the simulation do not activate this process (like in tungsten~\cite{hodille_2025}).
Finally, isotope exchange through ballistic detrapping is similar to isotope exchange through thermal detrapping, except that the detrapping is induced by an incoming plasma particle. 
Hence, it happens only in the ion implantation range, affecting only marginally the tritium recovery when D/T migrates deep into the bulk. In the future, isotope exchange via ballistic detrapping may be important for thin boron deposits.  

Equations \ref{eq:mobile-hydrogen}, \ref{eq:trapped-hydrogen}, \ref{eq:arrhenius-trapping}, and \ref{eq:arrhenius-detrapping} are solved in cartesian coordinates in 1D. Surface physics are accounted for via boundary conditions. HISP presently simulates single materials, where $\Gamma_1$ is the plasma facing simulation boundary, and $\Gamma_2$ is the cooling surface.

The value of the mobile concentrations at $\Gamma_1$ is set to zero when surface recombination is assumed fast.
The value of the material temperature at $\Gamma_1$ and $\Gamma_2$ is also imposed on the surfaces. 
These values can be set to any arbitrary function $f$ of spatial coordinates and time $t$. 
At $\Gamma_2$, HISP sets a no-flux boundary condition:
\begin{equation}
    \label{eq:neumann}
    J_D \cdot \boldsymbol{n} = -D \nabla c_{\mathrm{D}} \cdot \boldsymbol{n} = 0
\end{equation}
where $\boldsymbol{n}$ is the normal vector to boundary $\Gamma_2$ and the same equation is followed for tritium. 
Finally, when recombination is not instantaneous, a recombination flux boundary condition is assumed at $\Gamma_1$ for \gls{dd} with

\begin{equation}
    \label{eq:recomb-flux}
    J_\mathrm{D} \cdot \boldsymbol{n} = -D \nabla c_{\mathrm{D}} \cdot \boldsymbol{n} = -2 \ K_r c_{\mathrm{D}}^2
\end{equation}
where $K_r$ [\si{m^4.s^-1}] is the recombination coefficient. 
The same equation follows for TT recombination. 
For \gls{dt} recombination, Equation \ref{eq:dt-recombo} is used: 
\begin{equation}
    \label{eq:dt-recombo}
    J_{\mathrm{DT}} \cdot \boldsymbol{n} = - \ K_r c_{\mathrm{D}} c_{\mathrm{T}} .
\end{equation}
This model concatenates the adsorption, absorption, and recombination processes happening on the material surface~\cite{PICK1985}, hence a overall process of second order.
In HISP, boundary conditions are associated with a model's material, and described in detail in Section~\ref{sec:binning}.

\subsection{Plasma Data Handling}\label{sec:plasma-data-handling}
The Plasma Data Handling class creates, stores, and organizes every pulse type that is simulated in HISP. For this study on ITER, HISP uses several pulse types with the following specifications and data sources:  


\begin{itemize}
    \item ITER Q=10 Baseline Pulse
    \begin{itemize}
        \item \gls{dt} pulses run with Q=10 and full current.
        \item Data from 3 SOLPS cases with different far-SOL extrapolations from~\cite{KHAN2019}.
    \end{itemize}
    \item Low Power Deuterium (DD)
    \begin{itemize}
        \item Pure deuterium plasmas with particle and heat fluxes reduced by a factor of 5 compared to \gls{dt}.
    \end{itemize}
    \item Glow Discharge (\gls{gdc}) \cite{glow-1}
    \begin{itemize}
        \item 24-hour deuterium glow discharge plasma characterized by a uniform $D_2^+$ ion flux on the \gls{fw} and upper divertor surfaces with impact energy of \SI{500}{V}. 
        \item Assume dissociation upon impact, creating an ion flux of \SI{0.4}{A/m^2} or \SI{2.5e18}{D^+/m^2}.
    \end{itemize}


    \item Baking (BAKE)
    \begin{itemize}
        \item One-week period of raised reactor temperature (in ITER, to \SI{220}{^{\circ} C}). 
        \item No plasma. 
    \end{itemize}
\end{itemize}

For each pulse, users are required to specify ramp up, ramp down, steady state, and between pulse plasma times. HISP represents the ramp up and ramp down as a simplified linear model of particle flux and temperature. HISP has the capability also to simulate \gls{icwc} and tokamak plasma with \glspl{risp}, of which the results will follow in future work. The pulses presented here can be arranged to create a scenario, which is described by the Scenario class in HISP. 

\subsection{Operation scenarios}\label{sec:scenario}

\begin{figure}[h!]
    \centering
    \includegraphics[width=\linewidth]{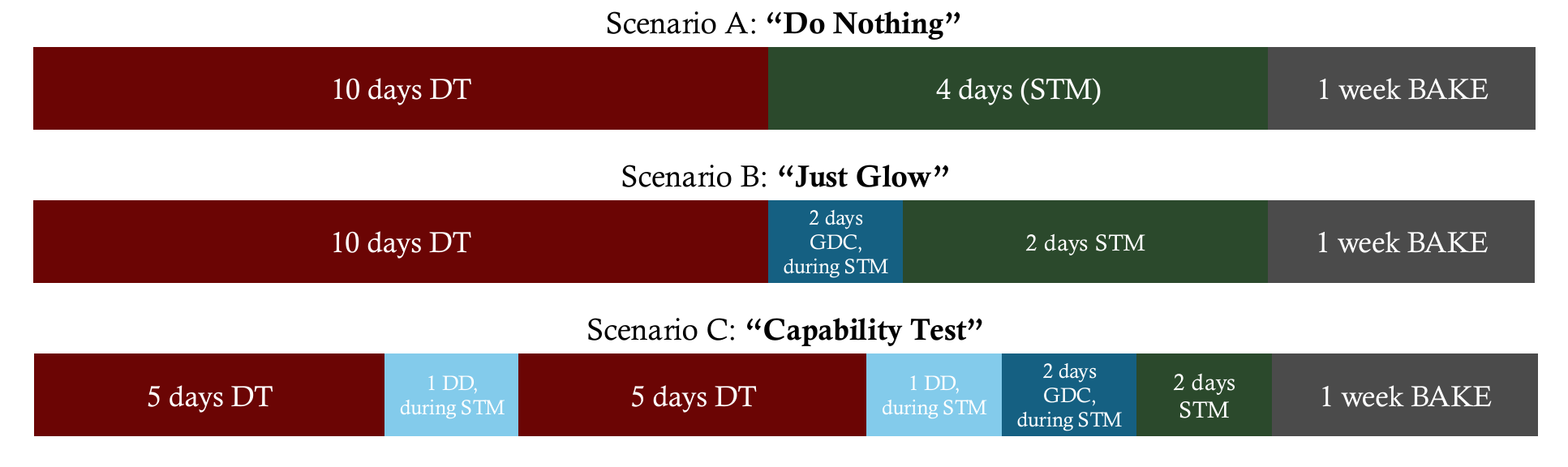}
    \caption{The three two-week scenarios tested in ITER, composed of pulses described in Section~\ref{sec:plasma-data-handling}.}
    \label{fig:scenarios}
\end{figure}

In this paper, three scenarios are tested on ITER, described in Figure~\ref{fig:scenarios}.
Plasma pulses are concatenated (including ramp up and ramp down), meaning 1 day of 13 \gls{dt} or \gls{dd} pulses as described in Figure~\ref{fig:cycle-timeline} is simulated as 1 longer \gls{dt} or \gls{dd} pulse.
This has been shown to have a negligible impact on inventory estimates \cite{delaportemathurin:tel-04004369, hodille-tungsten}.
For the purposes of demonstration, the results included in this paper also have shortened wait times between pulses, meaning each scenario does not exactly fit into 14 days. 

\subsection{Binning}\label{sec:binning}
\begin{figure}[h!]
     \centering
     \begin{subfigure}[b]{0.45\textwidth}
        \centering
        \includegraphics[width=0.7\textwidth]{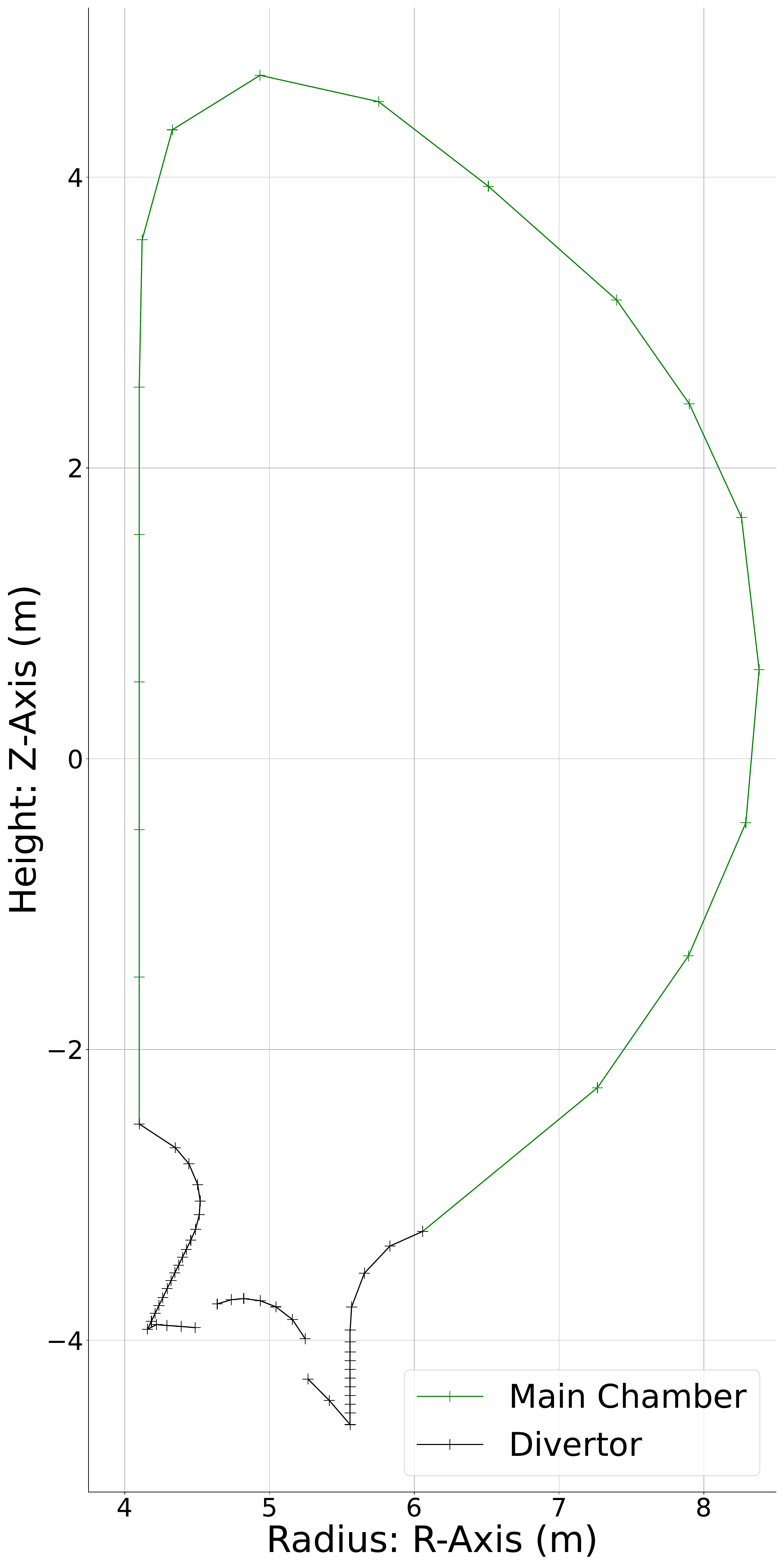}
        \caption{Binned ITER Cross Section.}
    \end{subfigure}
     \hfill
     \begin{subfigure}[b]{0.45\textwidth}
         \centering
         \includegraphics[width=1\textwidth]{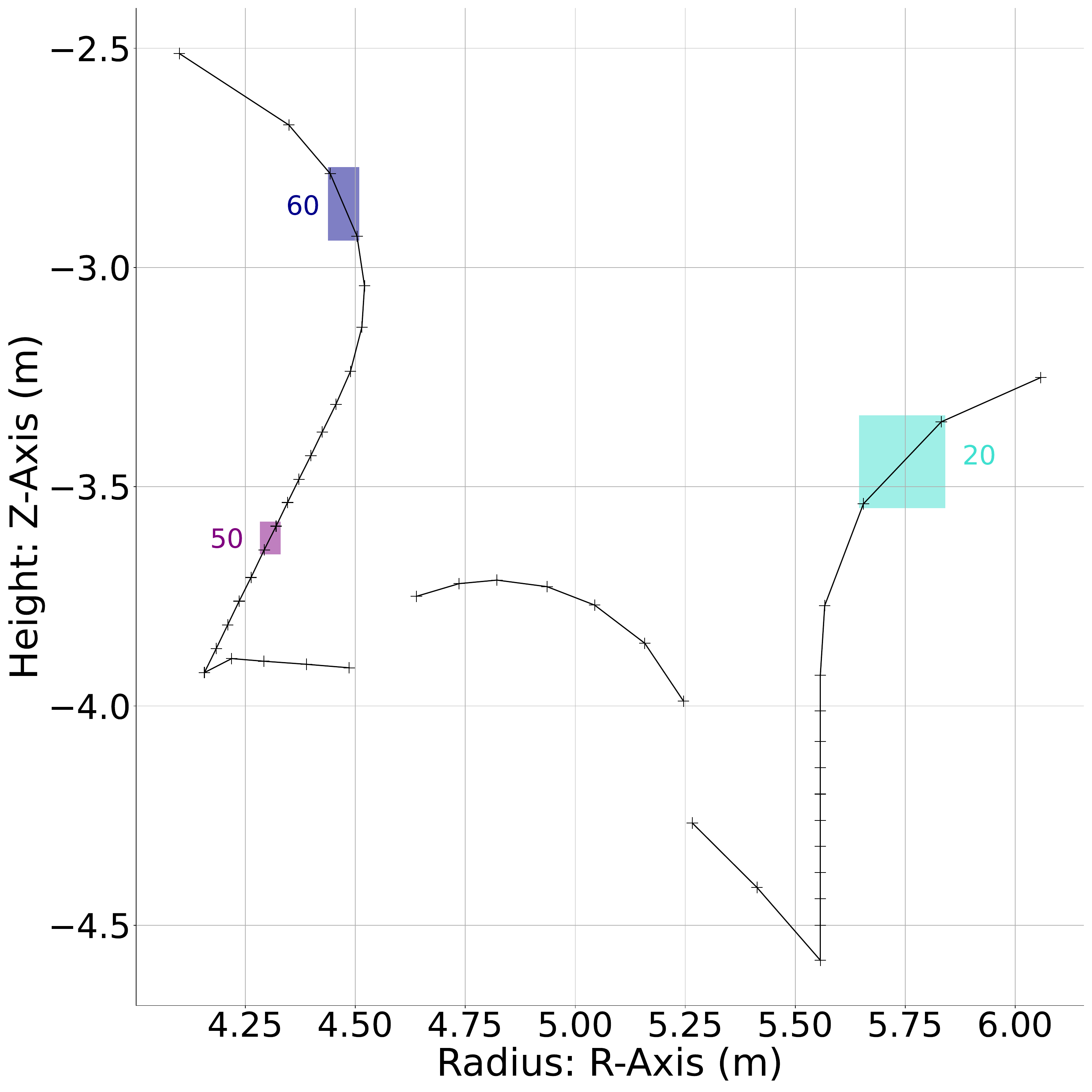}
         \caption{Divertor Bins, with highlights identifying bins detailed in Section~\ref{sec:results}.}
         \label{fig:div-bins}
     \end{subfigure}
    \caption{ITER PFC surfaces represented by HISP bins, here plotted as segments on the poloidal plane. The \gls{fw} is partitioned into 18 bins (each with sub-bins described in the following sub-sections), and the divertor into 44 bins. The total bin count, including \gls{fw} sub-bins, is 94.}
    \label{fig:iter-bins}
\end{figure}

HISP processes plasma data with a ``binning'' methodology, in which a reactor's PFCs are broken into representative surfaces for which independent one-dimensional hydrogen transport simulations are run. In this study, HISP partitions ITER's \gls{fw} and divertor components into 97 bins, plotted as segments on a poloidal cross-section in Figure~\ref{fig:iter-bins}. This binning framework allows HISP to spatially average particle and heat fluxes from plasma code sources into a description readable by FESTIM. 

\begin{figure}[hbt!]
    \centering
    \includegraphics[width=\linewidth]{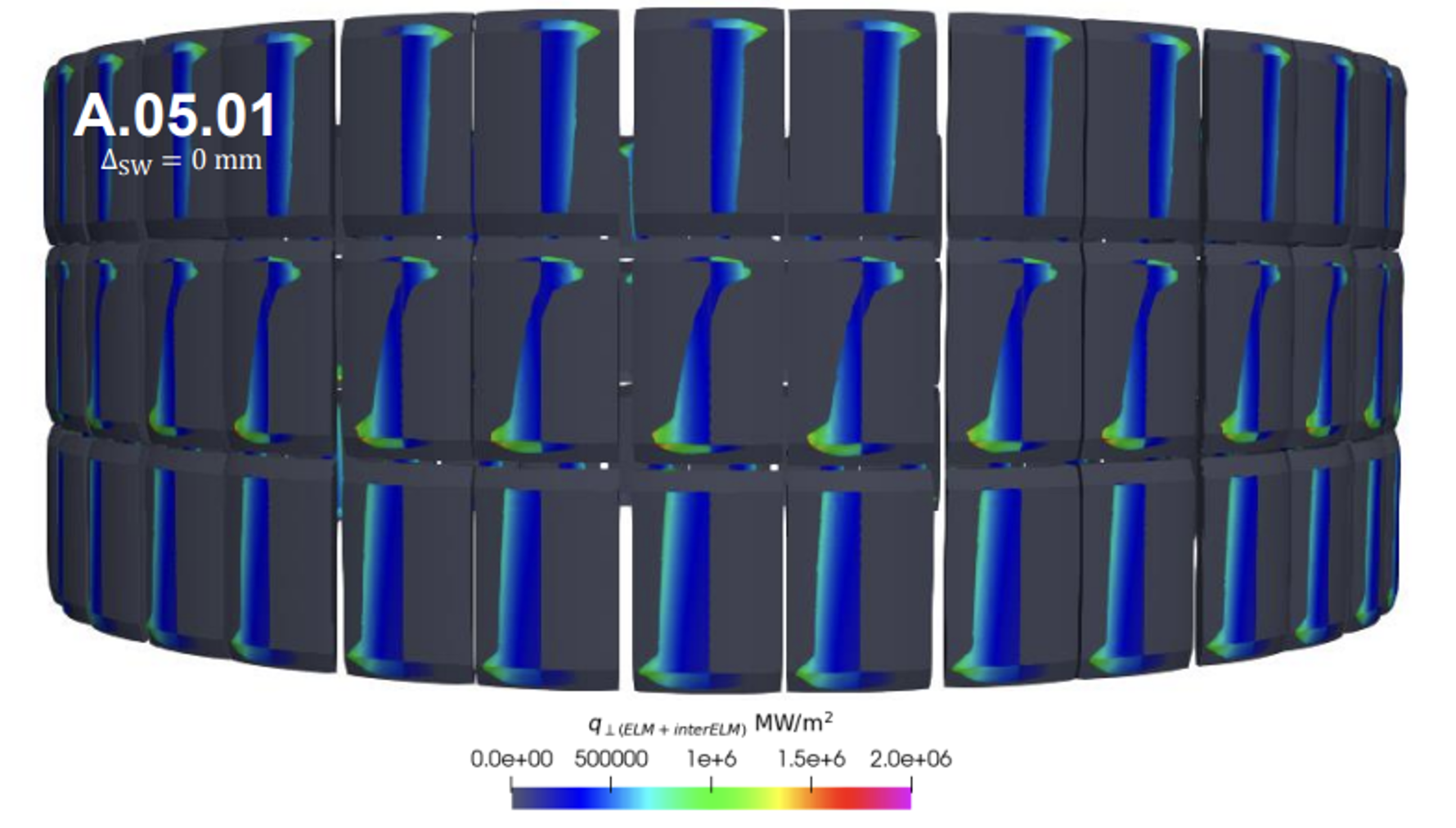}
    \caption{Illustration of heat load maps on \gls{fw} panels (from bottom to top: row 3, 4, and 5) used for assigning wetted (colored) and shadowed (gray) surface areas and averaged loads to the 1D bin simulations of HISP. The maps result from SMITER analysis by F. Fernández-Marina of a plasma baseline scenario with $\Delta$rSEP $=$ \SI{10}{cm}, using the panel shaping of the Be wall. Wetted areas receive both atom and ion fluxes, while shadowed areas receive only atom fluxes.}
    \label{fig:magorientation}
\end{figure}

The ITER \gls{fw} design has evolved toward shaped \gls{fw} panels to accommodate the actual plasma heat loads on the \gls{fw} and to reduce heat loads on leading edges down to acceptable levels \cite{iter-fw-shaping}. For a given tokamak plasma scenario, this results in wetted and shadowed areas, caused by magnetic field topology (see Figure~\ref{fig:magorientation}). This impacts the particle fluxes relevant for HISP simulations; wetted areas receive both ion and atom fluxes, while shadowed areas receive only atom fluxes. The HISP binning method accounts for these ``wetted" and ``shadowed" areas on \gls{fw} bins. For panels with significantly non-uniform ion heat loads, a high-wetted area and a low-wetted area can be defined. Hence, for the \gls{fw} representation in HISP, each parent bin (of which there are 18, one for each \gls{fw} row) obtains two to three sub-bins (one always reserved as a shadowed bin) determined by the heat profiles provided by SMITER simulations. SMITER is a magnetic field line tracing code that has been used to produce magnetic heat flux profiles for ITER \cite{KOS20191796}.


From the SMITER simulations, surface heat flux distributions at the plasma wetted area were obtained. In the case of non-uniform ion heat loads, two distinct peaks were observed in the heat distribution. In the case of more uniform ion heat loads, one distinct peak was observed. These distributions indicated whether two or three sub-bins would be used for each parent \gls{fw} bin. 
For parent bins with a high-wetted and a low-wetted sub-bin, total ion heat $H_{\mathrm{T}}$ [\si{W.m^{-2}}] and particle fluxes $F_{\mathrm{T}}$ [\si{m^{-2}.s^{-1}}], obtained from edge codes which assume toroidal symmetry, are scaled to the corresponding wetted surface area, respectively $a_1$ and $a_2$ [\si{m^2}]. 
The fraction of ion heat load on the high wetted area is defined as 

\begin{equation}\label{eq:f_high}
    f = \frac{H_1 a_1}{H_{\mathrm{T}} a_{\mathrm{T}}} = \frac{h_1 a_1}{h_1 a_1 + h_2 a_2}
\end{equation}

and that of the low wetted area as

\begin{equation}\label{eq:f_low}
    1-f = \frac{H_2 a_2}{H_{\mathrm{T}} a_{\mathrm{T}}} = \frac{h_2 a_2}{h_1 a_1 + h_2 a_2}
\end{equation}

where $h_1$ and $h_2$ [\si{W.m^{-2}}] are the average ion heat load from SMITER at $a_1$ and $a_2$, respectively, $H_1$ and $H_2$ [\si{W.m^{-2}}] are the heat loads imposed at the respective sub-bins in the HISP simulation, and $a_{\mathrm{T}}$ [\si{m^2}] is the total parent bin surface area which, with exception of rows with ports, is estimated as $a_{\mathrm{T}} = 2 \pi Rh$, with $R$ [\si{m}] the average radius of the parent bin segment and $h$ [\si{m}] its poloidal length. We again remember that each parent bin segment in the present HISP implementation corresponds to a \gls{fw} row. Equations \ref{eq:f_high} and \ref{eq:f_low} are solved for $H_1$ and $H_2$, which become inputs for HISP.

The same fraction is used to scale the ion flux, where $F_1$ [\si{m^{-2}.s^{-1}}] is the ion flux in the high wetted area, $F_2$ [\si{m^{-2}.s^{-1}}] is the ion flux in the low wetted area, and $F_{\mathrm{T}}$ is the total ion flux to that row provided by plasma code output. 
\begin{equation}
    F_1 = \frac{f F_{\mathrm{T}} a_{\mathrm{T}}}{a_1} 
\end{equation}
\begin{equation}
    F_2 = \frac{(1-f) F_{\mathrm{T}} a_{\mathrm{T}}}{a_2} 
\end{equation}
Atom and radiation fluxes are assumed uniform over $a_{\mathrm{T}}$. The surface area $a_0$ corresponding to the shadowed area sees atoms and radiation only, and is found using $a_{\mathrm{T}} = a_0 + a_1 + a_2$.

Finally, to accommodate for \gls{fw} rows equipped with ports, an additional shadowed sub-bin can be defined. 
Additional bins to represent the stainless steel diagnostic (DFW) surface can also be added in HISP, but are not simulated for this analysis. 
A complete list of \gls{fw} bins and sub-bins can be found in~\ref{appendix-binning}.

\subsection{HISP Physics Assumptions}\label{sec:assumptions}

HISP makes a number of physics assumptions. First, HISP assumes a completely empty \gls{fw} and divertor--each modeled bin starts with an inventory of zero. In reality, when boron layers are deposited, they will already contain a significant amount of fuel, potentially altering tritium inventory results. Similarly, before \gls{dt} operations starts in ITER, B and W bins will have been exposed to significant H and D flux. Second, the boron layer thicknesses are assumed to be constant; in other words, the rate of deposition and the rate of erosion for ITER's boron layers are assumed to be the same. Third, no fuel exchange between the thin boron layer and the W substrate is assumed. In absence of solubility data for B deposit layers, at present no meaningful boundary condition at the interface could be set. Next, the HISP boron model contains no D/T atom swapping due to the lack of isotopic exchange data available on the topic (see Section~\ref{sec:HTMFES}). Also for W, isotope swapping is not activated in the present analysis. 
Finally, HISP also assumes a simplified linear ramp up and ramp down model in plasma pulses. In reality, ITER's ramp up and ramp down will be nonlinear \cite{Kim_2018}. 

\section{FESTIM Model Parametrization}
While FESTIM has the ability to model any material with sufficient data, in this study, all bins and sub-bins are one of two models: the tungsten (W) model, or the boron (B) model. HISP was tested on the SS316 model to describe ports, but results for SS316 lie outside the scope of this paper. 

\subsection{Tungsten Model}\label{sec:tungsten}
In ITER, the W model is used for the majority of the wetted \gls{fw}. 
Although ITER's entire divertor armour is also made of W, areas with B co-deposits are represented in this study by the B model.
Thus, only the divertor strike points and vertical plates are tested with the W model.
The tungsten model uses material properties summarized in Table~\ref{tab:tungsten} and includes the two traps described in Table~\ref{tab:tungstentraps}. 

\begin{table}[hbt!]
    \centering
    \begin{tabular}{lr}
        Parameter & W Value \\
        \hline
        \\
         Metal Density & \SI{6.34e28}{m^{-3}}\\
         Interstitial Distance & \SI{1.12e-10}{m} \\
         Interstitial sites per metal atom & 6 \\
         Diffusion Coefficient $D_0$ & \SI{2.06e-7}{m^2.s^{-1}} \\ 
         Diffusion Energy $E_D$ & \SI{0.28}{eV} \\ 
    \end{tabular}
    \caption{The material properties used for the tungsten model. Metal density and interstitial properties from \cite{hodille-tungsten}. Diffusion properties from \cite{holzner}.}
    \label{tab:tungsten}
\end{table}

\begin{table*}[hbt!]
    \centering
    \begin{tabular}{>{\raggedright\arraybackslash}p{1.5cm}>{\raggedleft\arraybackslash}p{3cm}>{\raggedleft\arraybackslash}p{3cm}>{\raggedleft\arraybackslash}p{3cm}}
        Trap & Detrapping Energy $E_\mathrm{dt}$ (eV) & Trapping Energy $E_\mathrm{t}$ (eV) & Atomic Fraction (at.fr.)\\
        \hline
        Trap 1 & 0.85 & 0.28 & \num{e-4}\\
        Trap 2 & 1.00& 0.28& \num{e-4}\\
    \end{tabular}
    \caption{Traps for the tungsten model from \cite{hodille-tungsten}.}
    \label{tab:tungstentraps}
\end{table*}
	
The 1D HISP tungsten model includes two boundary conditions. On the plasma-facing side of each tungsten bin is a recombination boundary condition calculated from a simplified surface kinetic model. Following the model used in \cite{MONTUPETLEBLOND2021101062}, the recombination coefficient $K_r(T)$ is expressed as: 

\begin{equation}
    K_r(T) = \frac{\nu_{bs}^2(T)\nu_{d}(T)}{\nu_{sb}^2(T)}
\end{equation}

where $\nu_{bs}$ is the resurfacing rate, $\nu_{d}$ the desorption rate, and $\nu_{sb}$ the absorption rate, all in [\si{s^{-1}}]. These rates were calculated using values in~\cite{Hodille_2024} to yield a recombination coefficient of \SI{7.94e-17}{m^4.s^{-1}} and a recombination energy of \SI{-2.0}{eV} for tungsten. At the cooling surface of the tungsten model, a no-flux boundary condition is assumed for reasons discussed in Section \ref{sec:assumptions}. 



    

    


In ITER, the temperature profiles for the tungsten models differ depending on their location in the reactor. \gls{fw} tungsten models share an interface with cooling materials (either Cu or CuZrZr), while divertor tungsten models do not. In this case, the temperature drop across the \gls{fw} tungsten slabs ($\Delta T_W$) must be calculated using 
\begin{equation}
    \Delta T_W = \frac{q_{f} d_w}{k_W}
\end{equation}
where $q_f$ [\si{W.m^{-2}}] is the heat flux at the plasma side of the tungsten slab, $d_w$ [\si{m}] is the thickness of the tungsten slab, and $k_W = $\SI{170}{W.m^{-1}K^{-1}} \cite{FLADISCHER2019178373} is the thermal conductivity of tungsten. The temperature drop across the copper slab, $\Delta T_{Cu}$, must also be calculated, using 
\begin{equation}
    \Delta T_{Cu} = \frac{q_f D_{Cu}}{k_{Cu}}
\end{equation}
with the thickness of the copper slab $D_{Cu}$ [\si{m}] and the thermal conductivity of copper $k_{Cu} = $ \SI{400}{W.m^{-1}K^{-1}} \cite{CHO2018197}. The temperature drop at the copper-water interface can be calculated using 
\begin{equation}
    \Delta_{T_{int}} = \frac{q_f}{h_{Cu}}
\end{equation}
where $h_{Cu} = $ \SI{10,000}{W.m^{-2}K^{-1}} \cite{Caruso22012015} is the heat transfer coefficient from copper to water. Finally, we can calculate the temperature at the tungsten-copper interface $T_{int}$, 

\begin{equation}
    T_{int} = T_c + \Delta T_{int} + \Delta T_{Cu}
\end{equation}
with the cooling temperature set at $T_c = $ \SI{343}{K}. The temperature at the plasma-facing surface of the tungsten slab $T_s$ is then

\begin{equation}
    T_s = T_{int} + \Delta T_{W} .
\end{equation}

For tungsten monoblocks in the divertor, the temperature of the plasma-facing surface is calculated as
\begin{equation}\label{eq:remi-tungsten-surface}
    T_s = \num{1.1e-4} q_f + T_c 
\end{equation}
from  \cite{delaporte-mathurin-iter-study}. For the rear side of divertor monoblocks, a similar model is used, 
\begin{equation}\label{eq:rear-simple}
    T_r = m_{r} q_f + T_c
\end{equation}
in which $m_r$ is a slope calculated from \cite{hodille-tungsten}, where tungsten monoblocks were exposed to three different heat fluxes and their resulting surface and rear temperatures were calculated. Using a linear regression model, the slope $m_r$ was calculated for use in our monoblock temperature model. 

Given the surface and rear temperatures of a tungsten slab in the \gls{fw} or a tungsten monoblock in the divertor, a linear temperature model is enforced throughout the tungsten.  

\subsection{Boron Model}\label{sec:boron}

Boron layers in ITER are deposited atop tungsten PFCs through boronization assisted by \gls{gdc} \cite{WAUTERS2025101891}. In subsequent plasma operations, the layers erode swiftly from wetted areas, and a significant fraction redeposits in the divertor \cite{SCHMID2024101789}. In our model, erosion and migration are not resolved, and tritium retention is only modeled in the boron co-deposited layers, ignoring the W substrate, represented by 1D boron bins. For ITER PFC representation in this study, to represent boronization-erosion-redeposition, B bins are located at some shadowed and low-wetted \gls{fw} bins, as well as the dome, outer plate, inner plate, outer curved strike point, and inner curved strike point of the divertor. The total amount of boron in the simulation corresponds to \SI{1.26}{kg}, or 14.6 boronizations (given each boronization introduces \SI{86}{g} of boron to the reactor). A complete list of ITER bins by materials can be viewed in~\ref{appendix-binning}. 
The boron model uses material properties summarized in Table~\ref{tab:boron} and includes the four traps described in Table~\ref{tab:borontraps}. 
This model is obtained by simulating deuterium TDS spectra from boron thin films after ion implantation~\cite{OYA2004}. See \ref{app:Boron} for the TDS model validation.
The parameters summarized in Tables \ref{tab:boron} and \ref{tab:borontraps} were also used to estimate the efficiency of baking for tritium recovery from boron deposits~\cite{PITTS2025101854}. 

\begin{table}[hbt!]
    \centering
    \begin{tabular}{lr}
        Parameter & B Value \\
        \hline
        \\
         Metal Density & \SI{1.34e29}{m^{-3}}\\
         Interstitial Distance & \SI{8.00e-10 }{m} \\
         Interstitial sites per metal atom & 1 \\
         Diffusion Coefficient $D_0$ & \SI{1.07e-6}{m^2/s} \\ 
         Diffusion Energy $E_D$ & \SI{0.30}{eV} \\ 
    \end{tabular}
    \caption{Material properties used for the boron model \cite{PITTS2025101854}.}
    \label{tab:boron}
\end{table}

\begin{table*}[hbt!]
    \centering
    \begin{tabular}{>{\raggedright\arraybackslash}p{1.5cm}>{\raggedleft\arraybackslash}p{3cm}>{\raggedleft\arraybackslash}p{3cm}>{\raggedleft\arraybackslash}p{3cm}}
        Trap & Detrapping Energy $E_\mathrm{dt}$ (eV) & Trapping Energy $E_\mathrm{t}$ (eV) & Atomic Fraction (at.fr.)\\
        \hline
        Trap 1 & 1.052 & 0.3 & \num{6.867e-1}\\
        Trap 2 & 1.199 & 0.3& \num{5.214e-1}\\
        Trap 3 & 1.389 & 0.3& \num{2.466e-1}\\
        Trap 4 & 1.589 & 0.3& \num{1.280e-1}\\
    \end{tabular}
    \caption{Traps for the boron model from \cite{PITTS2025101854}.}
    \label{tab:borontraps}
\end{table*}

The 1D HISP boron model is an interim approach (meaning simple boundary conditions), given the current lack of experimental data on tritium transport in boron. On the plasma-facing side of each boron bin is a Dirichlet boundary condition with mobile concentration $c_{\mathrm{m},i} = 0$. On the tungsten-facing side of the boron bin, a no-flux boundary condition is enforced. 

The temperature of a given boron layer is assumed to be homogeneous (due to the thinness of the boron bins) and is calculated based on the surface temperature of the tungsten beneath it (see Section \ref{sec:tungsten}). The boron temperature is found using
\begin{equation}\label{eq:borontemp}
    T_B = R_c \ q_f + T_s
\end{equation}
with $R_c = \SI{5e-4}{m^2K/W}$ the thermal contact resistance at the boron-tungsten interface, a value obtained for Be and C deposits on W in JET \cite{risp, GASPAR2016292}, while similar observations are made for boron containing deposits in WEST \cite{GERARDIN2024101783}. $q_f$ remains heat flux, and $T_s$ is the surface temperature of the tungsten slab beneath the boron layer as calculated from Equation~\ref{eq:remi-tungsten-surface}.

\section{Results}\label{sec:results}
\begin{figure}[h!]
     \centering
     \begin{subfigure}[b]{0.45\textwidth}
         \centering
         \includegraphics[width=\textwidth]{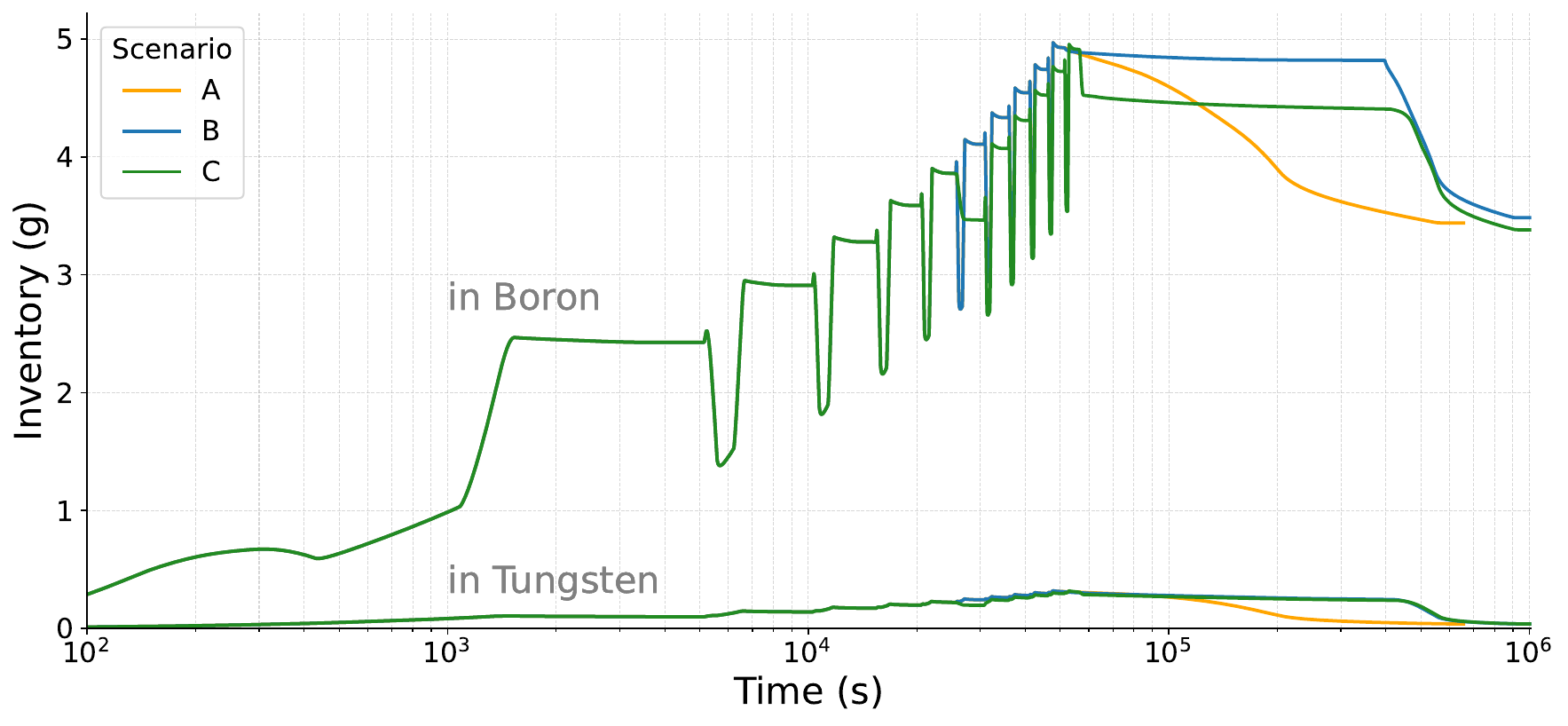}
         \caption{Wall.}
         \label{fig:comp-wall}
     \end{subfigure}
     \hfill
     \begin{subfigure}[b]{0.45\textwidth}
         \centering
         \includegraphics[width=\textwidth]{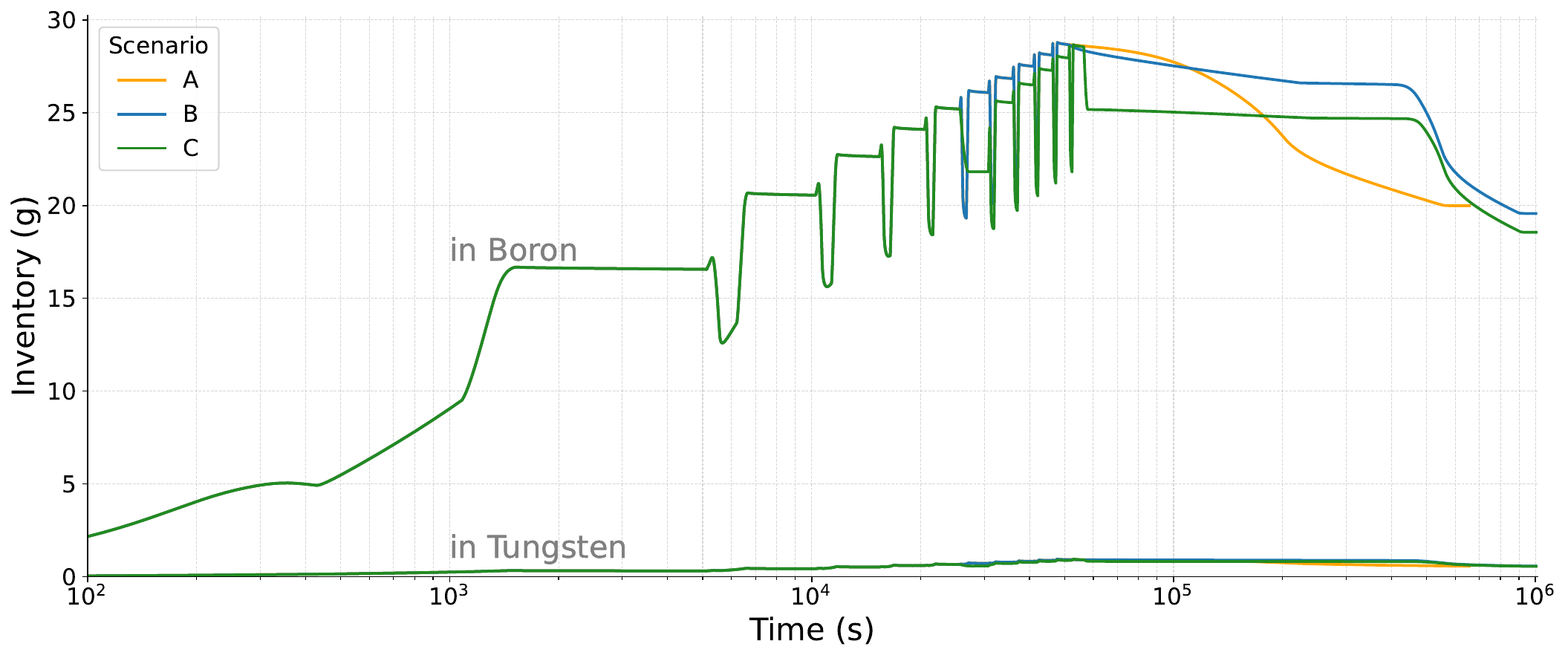}
         \caption{Divertor.}
         \label{fig:comp-div}
         
     \end{subfigure}
     
    \caption{Evolution of tritium inventory in the \gls{fw} and divertor at the end of scenarios A, B, and C.}
    \label{fig:total-10-pulses}
\end{figure}

While this paper is intended to focus on the HISP methodology, some results are included to demonstrate HISP's capabilities, and overall tritium inventory trends during \gls{dt} and cleaning operations. Our results suggest that all three removal scenarios are comparable in their reduction of tritium inventory (see Figure \ref{fig:total-10-pulses}), but Scenario C results in marginally less tritium inventory at its conclusion. 
Final inventory values for each scenario vary from each other by less than \SI{2}{\%} in the \gls{fw} and \SI{10}{\%} in the divertor.
In both the \gls{fw} and divertor, during all three scenarios, the boron inventories dominate tungsten inventories by an order of magnitude in \si{g}, even though the layers are thin. This indicates that most of the tritium inventory will be concentrated in boron deposits during operational campaigns. We simulate only 2 weeks of operation, and given that tritium inventory in W scales as the square root of time \cite{delaporte-mathurin-iter-study}, a campaign of 16 months (or about 34 two-week cycles) would increase W inventory by 5 to 6 times. 
We attribute high inventory in boron to boron's four traps with high atomic fractions as compared to tungsten's 2 traps with 3 orders of magnitude lower atomic fractions. W does have a super saturated layer with high trap concentration \cite{Hodille_2021} that is not represented in HISP due to the thinness of the layer. In fact, it is thinner than even the B layers, which allows B to dominate the inventory. 
Our results confirm previous studies \cite{Zuo_2023, 10.1063/1.5026415, Wang_2012} that investigate boron co-deposits holding significant portions of tritium inventory in plasma facing components.

\begin{figure}[h!]
     \centering
     \begin{subfigure}[b]{0.45\textwidth}
         \centering
         \includegraphics[width=\textwidth]{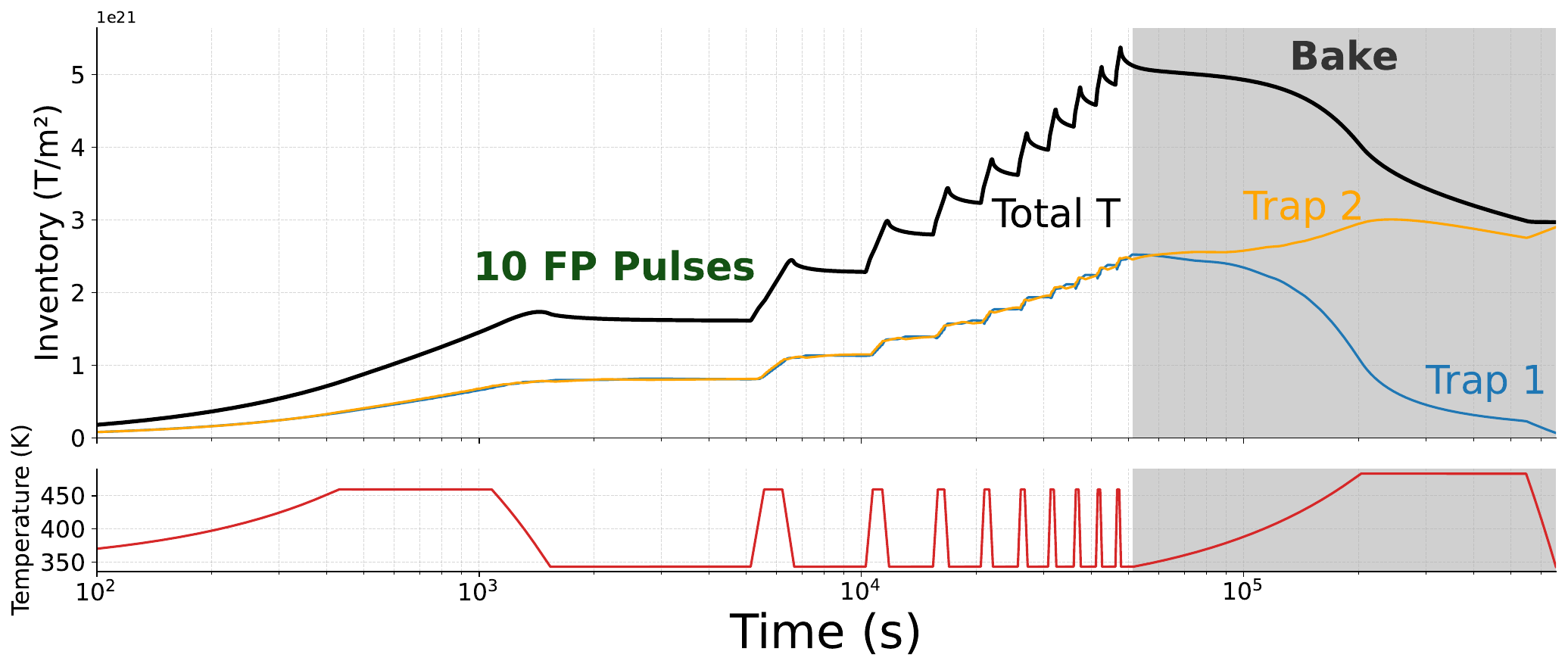}
         \caption{Full scenario.}
         \label{fig:W-scenA}
     \end{subfigure}
     \hfill
     \begin{subfigure}[b]{0.45\textwidth}
         \centering
         \includegraphics[width=\textwidth]{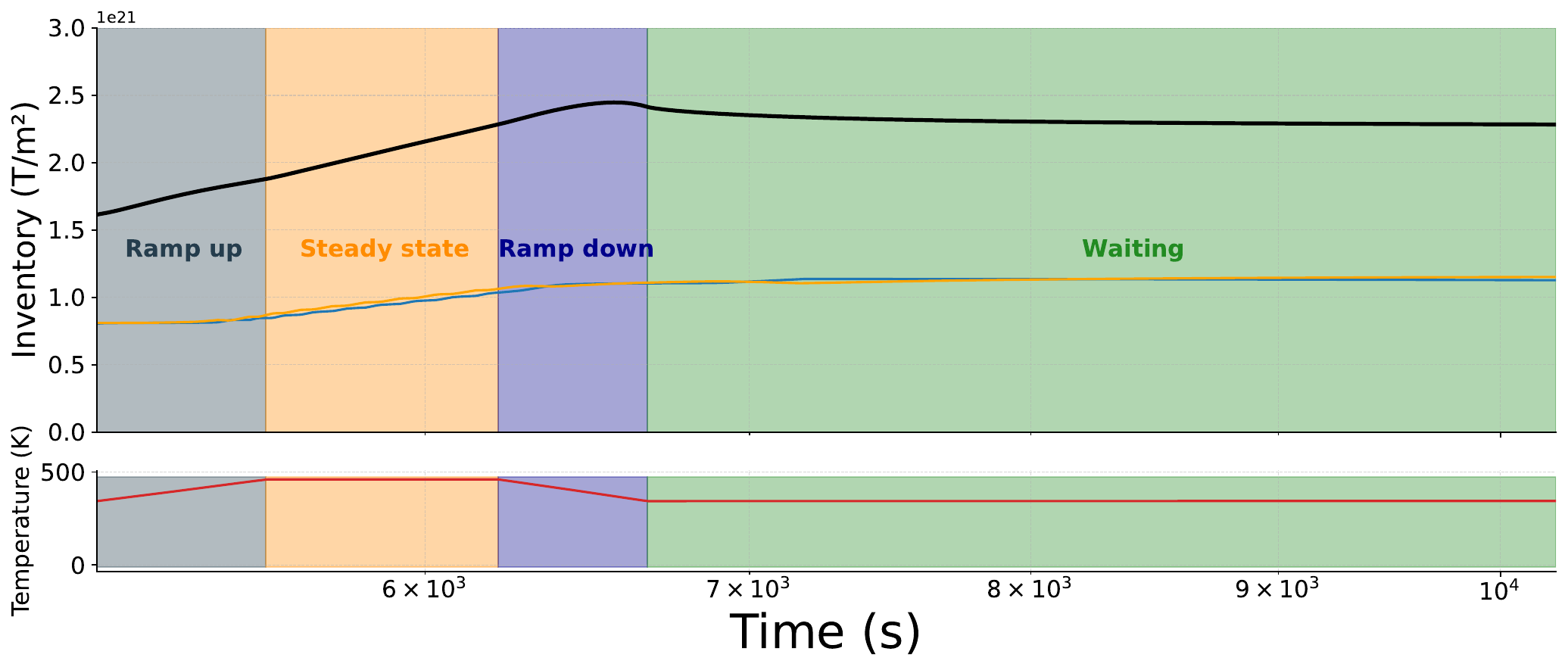}
         \caption{Third pulse.}
         \label{fig:W-scenA-zoom}
         
     \end{subfigure}
     
    \caption{Temporal evolution of the tritium inventories and temperature for Tungsten bin 50 (Scenario A). See Figure \ref{fig:div-bins} for bin location.}
    \label{fig:scenA}
\end{figure}

The trapped and mobile tritium inventories in tungsten are described in Figure \ref{fig:scenA}, where Figure \ref{fig:W-scenA} suggests that tungsten inventory is dominated by Trap 2, the highest energy trap. Baking demonstrates a trend as the highest efficiency for an individual cleaning technique, landing at a 87.99 $\%$ inventory decrease in the tungsten wall bins but dropping to a 39.73 $\%$ decrease in the tungsten divertor bins.  
Such a low baking efficiency for divertor monoblocks confirms the results of previous studies  \cite{Delaporte-Mathurin_2024,Delaporte-Mathurin_2021,Hodille_2017}
However, this study can surely be expanded in the future to include optimization of bake temperature for future reactors. 

The tritium inventory evolution throughout Scenario A for a bin located at the inner strike point is shown in Figure~\ref{fig:W-scenA}, with a close look at the behavior during a \gls{dt} pulse in Figure~\ref{fig:W-scenA-zoom}. 
Throughout ramp up and steady state, the inventory steadily increases as particle fluxes and heat loads increase. 
During the ramp down, the temperature is decreasing and the flux does not drop to zero immediately; the trapping/detrapping balance shifts in favor of trapping and tritium inventory increases. 
At some point along the ramp down, however, particle and heat fluxes become low enough to induce a decrease in inventory. 
Because we are using linearly simplified ramp ups and ramp downs, future work should implement realistic ramp up and ramp down representation to investigate the change in this behavior.



\begin{figure}
    \centering
    \includegraphics[width=0.5\textwidth]{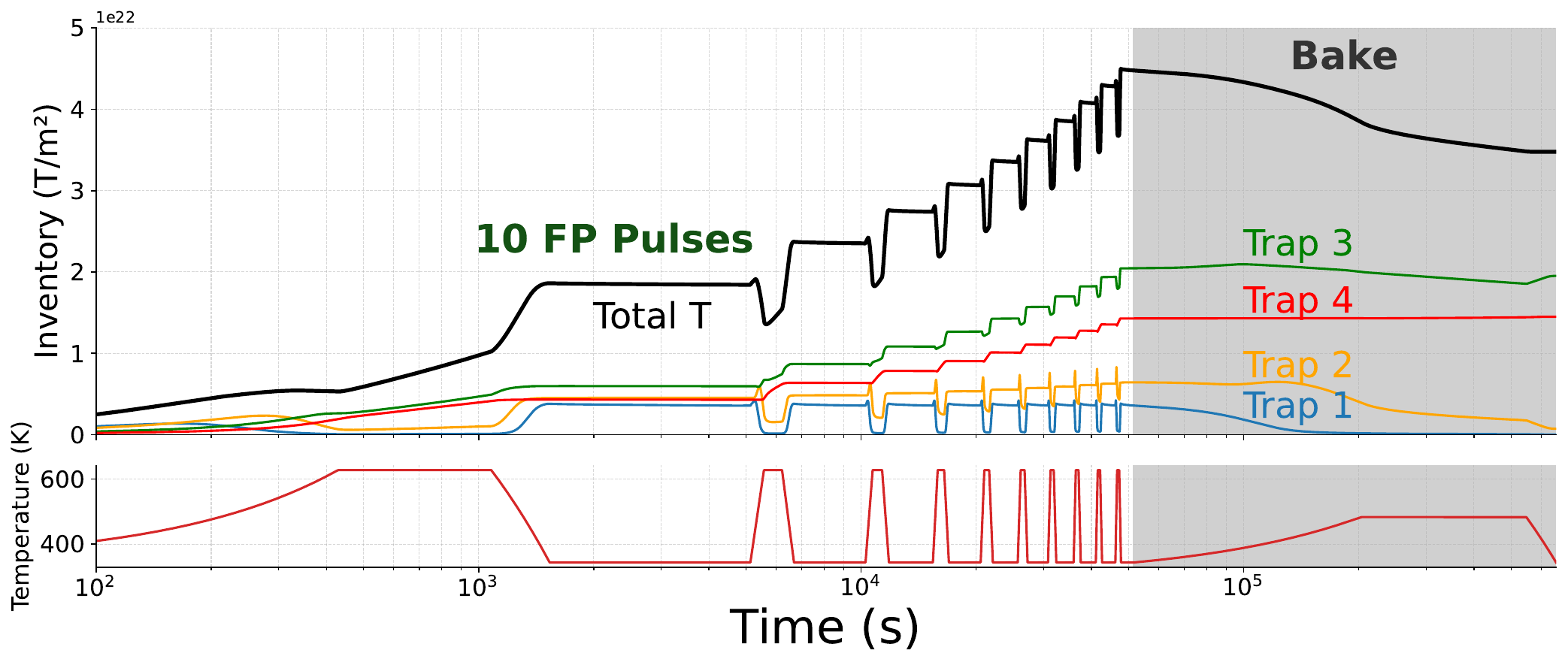}
    \caption{Temporal evolution of the tritium inventories and temperature for boron bin 60 (Scenario A). See Figure \ref{fig:div-bins} for bin location.}
    \label{fig:B_scenA}
\end{figure}

In boron, the bake temperature is between \SI{100}{K} to \SI{200}{K} lower than the loading temperatures, as seen in Figure \ref{fig:B_scenA}. Thus baking decreases inventory only by 30.80 $\%$ in B wall bins and by 30.59 $\%$ in B divertor bins.
The inventory of tritium trapped in Trap 4, the highest energy trap in boron, does not reduce due to baking. This indicates that higher temperatures may be needed to remove significant amounts of tritium from high energy traps in boron deposits. 
Given this reality, it may be difficult to remove significant amounts of tritium stored in ITER's \gls{fw}--where it is unlikely to achieve temperatures high enough to remove tritium. 

\begin{figure}[h!]
     \centering
     \begin{subfigure}[b]{0.45\textwidth}
         \centering
        \includegraphics[width=\textwidth]{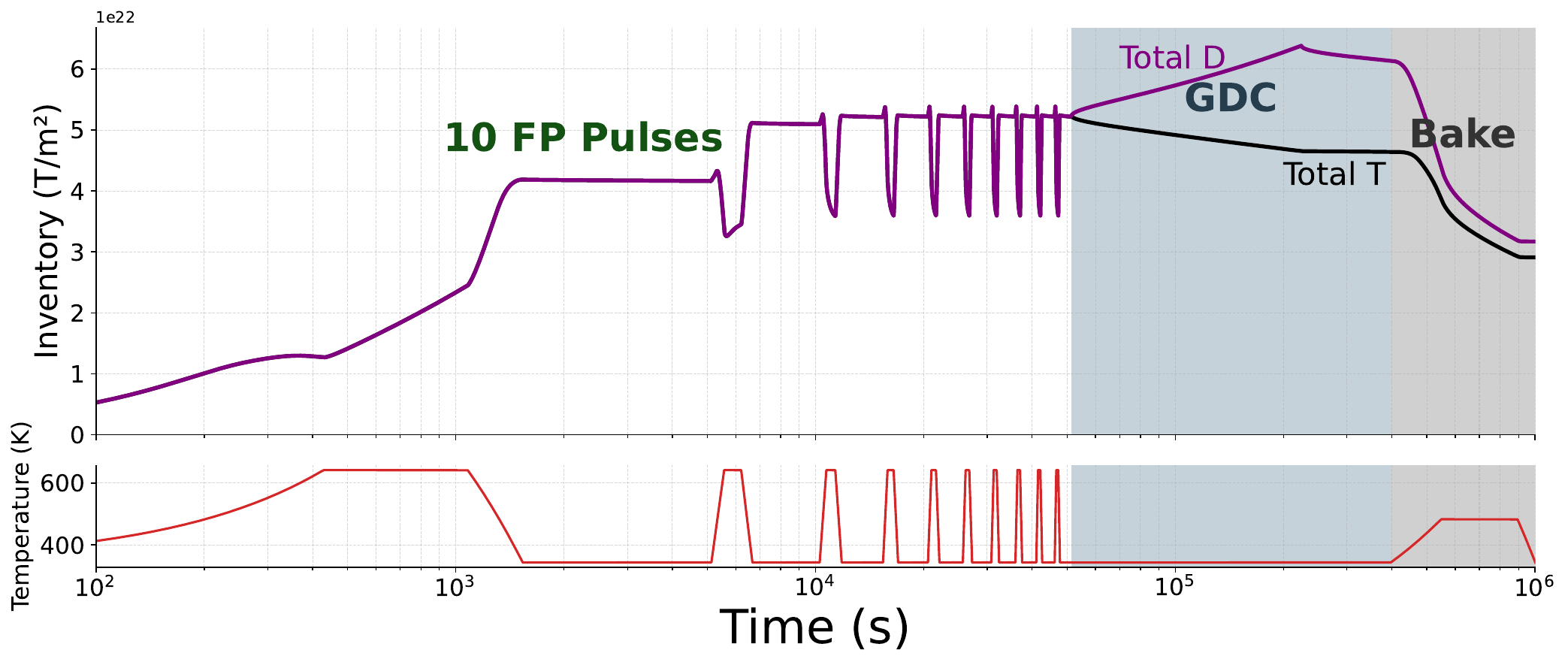}
         \caption{Full scenario.}
         \label{fig:scenB-B}
     \end{subfigure}
     \begin{subfigure}[b]{0.45\textwidth}
         \centering
         \includegraphics[width=\textwidth]{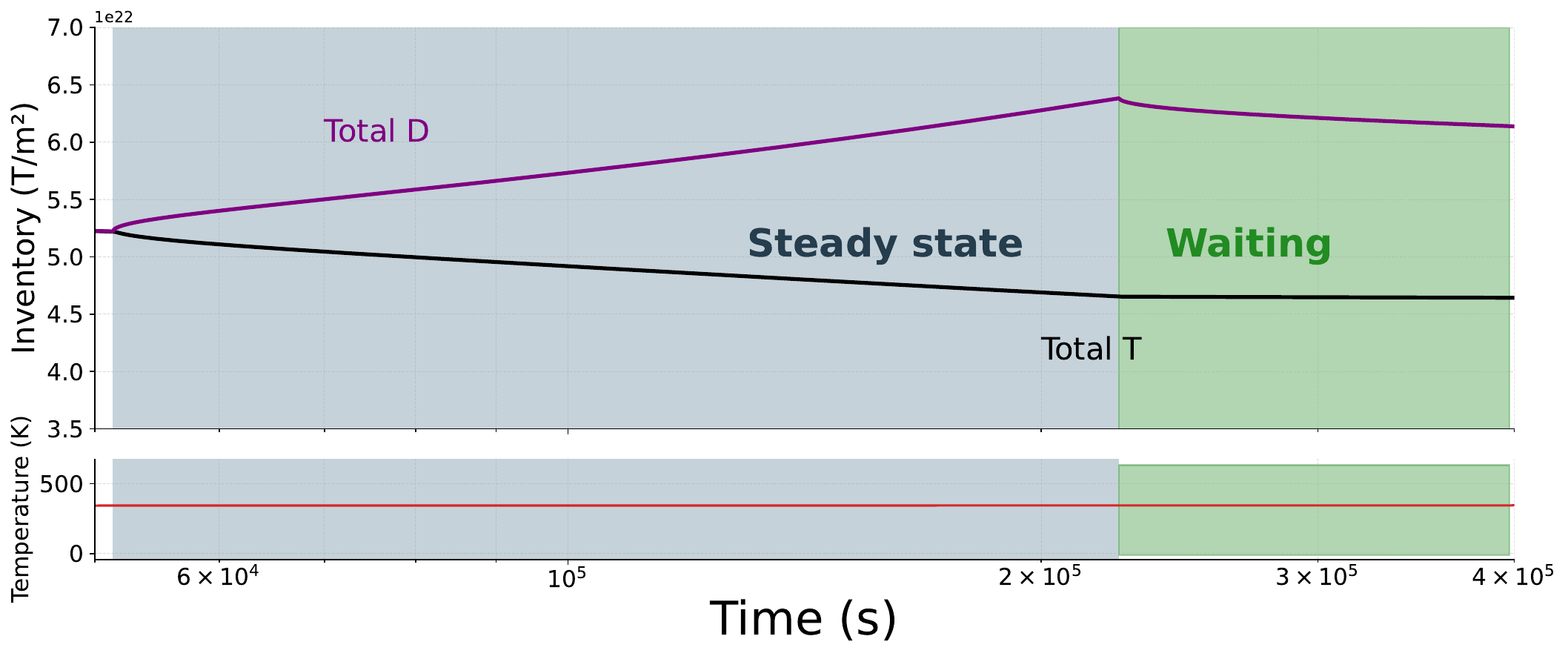}
         \caption{\gls{gdc} pulse.}
         \label{fig:scenB-B-zoom}
     \end{subfigure}
    \caption{Temporal evolution of the hydrogen inventories and temperature for boron bin 20 (Scenario B). See Figure~\ref{fig:div-bins} for bin location.}
    \label{fig:scenB}
\end{figure}

\gls{gdc} pulses demonstrate the multi-isotope transport capabilities in FESTIM. Figure \ref{fig:scenB} suggests the increase in deuterium inventory as a result of the pure \gls{gdc} pulse, and simultaneously the decrease in tritium inventory due to \gls{gdc}. \gls{gdc} pulses have only ion (no heat) fluxes, and thus a temperature of \SI{343}{K}, operating instead via indirect isotope exchange. As tritium inventory decreases from traps during the \gls{gdc} via isotopic exchange (see \cite{Puyang_2024, NAKAMURA2004163}), deuterium particles fill the traps made empty from tritium. This isotope exchange is driven by thermal detrapping (see Section~\ref{sec:HTMFES}), where a tritium atom exchanges with a deuterium atom through the detrapping of a tritium atom and subsequent deuterium atom trapping, resulting in the behavior demonstrated in Figure~\ref{fig:scenB-B-zoom}.


The \gls{gdc} efficiency in the tungsten \gls{fw} was calculated as a 23.01 $\%$ decrease in tritium inventory. 
HISP assumes that \gls{gdc} applies ion flux everywhere but the straight lower part of the vertical targets and horizontal plates of the divertor (both made of tungsten). 
This limits the ability of \gls{gdc} pulses to remove tritium inventory from the areas of the divertor that receive the highest heat and particle fluxes during full power operation. 
Thus \gls{gdc} contributes to no decrease of the tritium inventory for the tungsten divertor bins, but does create a 7.89 $\%$ decrease for boron divertor bins.

\begin{figure}[h!]
     \centering
    \includegraphics[width=0.5\textwidth]{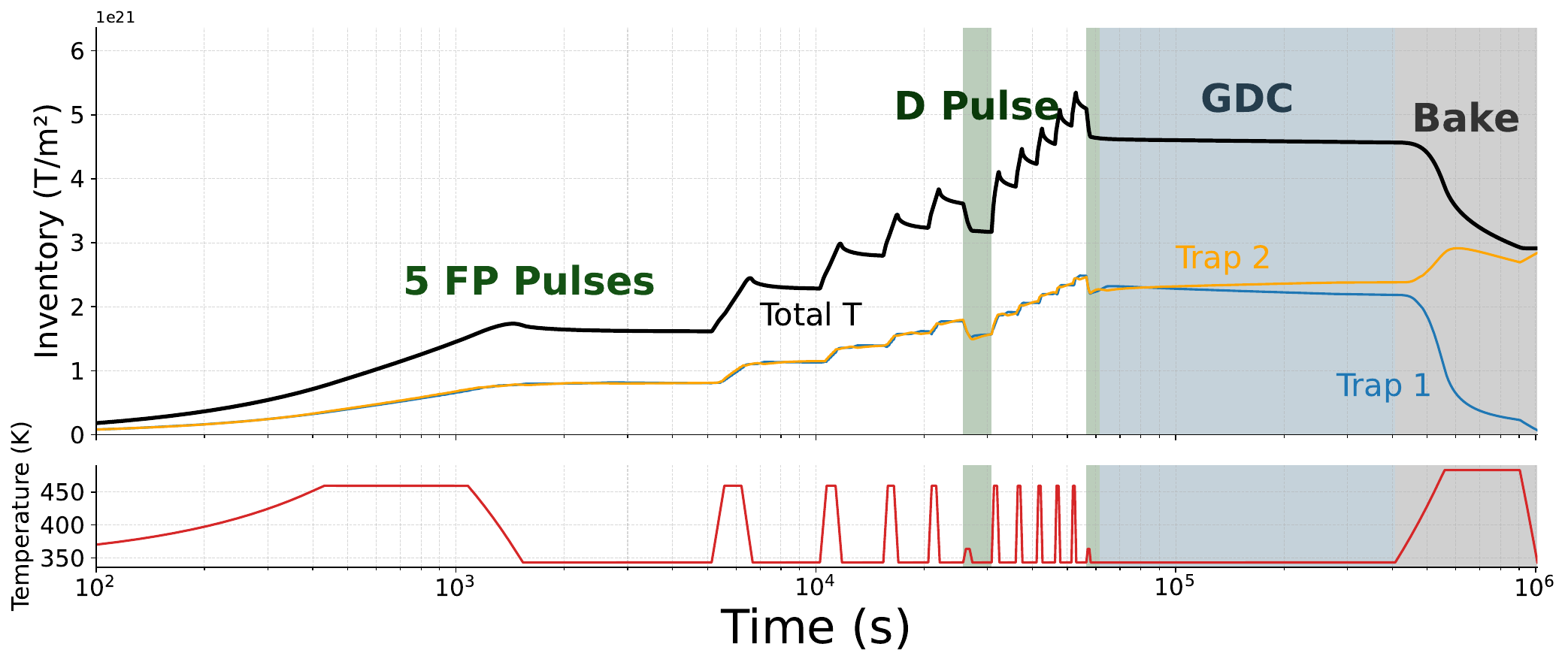}
    \caption{Temporal evolution of the hydrogen inventories and temperature for tungsten bin 50 (Scenario C). See Figure~\ref{fig:div-bins} for bin location.}
    \label{fig:scenC}
\end{figure}

Scenario C tested the cleaning efficiency of a low-power deuterium pulse between full power operation days, demonstrating in Figure \ref{fig:scenC}.
\gls{dd} pulses for the \gls{fw} decreased tritium inventory by about \SI{10}{\%} in both W and B. 
In the divertor, inventory was decreased by about \SI{13}{\%} for both W and B. 
In the case for bins that are not affected by \gls{gdc} pulses, as in Figure \ref{fig:scenC}, \gls{dd} pulses may be a good way to reduce tritium inventory before the longer period of baking.
Tritium inventory reduction is assumed to occur via isotopic exchange as in the \gls{gdc} pulse.
This small decrease throughout full power operation may be responsible for the marginally better performance of Scenario C following one cycle in Figure \ref{fig:total-10-pulses}. 
It is difficult to determine which scenario is best, due to reasons discussed in Section \ref{sec:discussion}, but as with Scenario A and Scenario B, Scenario C suggests that baking is the most effective tritium removal technique. 

\section{Discussion}\label{sec:discussion}
While important insights on tritium inventory build-up mitigation are obtained from this first of a kind reactor scale inventory analysis in PFCs using HISP, summarized in Figure \ref{fig:total-10-pulses}, it is yet hard to be conclusive about the results; there is a margin of unaccounted error that affects the results. Uncertainty values from plasma codes are not propagated through HISP simulations. 
Uncertainty calculations from hydrogen transport simulations, be it material parameters or boundary conditions, are also not collected in HISP. 
Given that final inventory values for each scenario vary marginally, these results contain unknown levels of error that could change the finding here of Scenario C as the optimal cleaning scenario. 

HISP currently exists at a proof-of-concept stage, and the results of this paper are intended to share approximate trends about tritium inventory with ITER as a case study. 
There are many ways for HISP to be improved, including propagating uncertainty throughout calculations as well as reducing file sizes. 
Data files created from boron bins lied anywhere from 1GB of data to 15GB of data for a single bin, making data parsing extremely difficult. 
Future work could include improvement in these areas, as well as multiphysics development of the model described in Section \ref{sec:conclusion}.

\section{Conclusion}\label{sec:conclusion}

The Hydrogen Inventory Simulations in PFCs (HISP) package was developed to simulate T inventory in magnetic confinement fusion devices, and is here used to assess the efficiency of various T removal techniques for ITER. 
HISP remains at a proof-of-concept stage.
Three operation scenarios were tested, each including \gls{dt} operations but varying in their inclusion of T removal techniques (either baking, \gls{gdc}, and/or \gls{dd} pulses).
Within individual scenarios, results suggested that baking was the most effective removal technique, decreasing T inventories in W $\approx$ \SI{88}{\%} in the \gls{fw} and $\approx$ \SI{40}{\%} in the divertor. 
For B, baking decreased T inventories by $\approx$ \SI{30}{\%} everywhere. 
\gls{gdc} and \gls{dd} T reduction in the \gls{fw} were \SI{23}{\%} and \SI{10}{\%} respectively; in the divertor, \SI{7}{\%} and \SI{13}{\%}. 
Ultimately, all three cleaning scenarios were comparable in their reduction of tritium inventory, with the scenario including most frequent cleaning as marginally optimal.
We anticipate that baking's dominance in T removal efficiency when compared to \gls{gdc} and \gls{dd} resulted in T inventories varying by less than \SI{2}{\%} in the \gls{fw} and \SI{10}{\%} in the divertor between scenarios.

HISP has potential to be improved with uncertainty propagation, as well as extension to multi-material and/or multi-dimensional studies. 
Comparison to experimental results in ITER, as well as devices like Tore-Supra/WEST and JET, would improve the accuracy of HISP estimates.
Optimization of cleaning scenarios, specifically bake temperatures, is another focus that future work could take. 
With these improvements, HISP can be used to satisfy regulatory requirements on tracing T inventory and optimizing cleaning methods for T removal. 

\glsaddall
\printglossary[title={Glossary of Terms}]

\section*{Acknowledgement}
This work was partially supported by the Ing Fellowship as granted by the Plasma Science and Fusion Center from the Massachusetts Institute of Technology. The views and opinions expressed herein do not necessarily reflect those of the ITER Organization. This work has also been carried out within the framework of the EUROfusion Consortium, funded by the European Union via the Euratom Research and Training Programme (Grant Agreement No 101052200 - EUROfusion). Views and opinions expressed are however those of the author(s) only and do not necessarily reflect those of the European Union or the European Commission. Neither the European Union nor the European Commission can be held responsible for them.

\section*{Data Availability}
Data used for the results in this paper can be found in the PFC tritium transport, \href{https://github.com/iterorganization/PFC-Tritium-Transport}{PFC-TT, repository} hosted on the ITER Organization Github page. 

\appendix
\section{ITER Bins}
\label{appendix-binning}

\begin{figure}[h!]
    \centering
    \includegraphics[width=\linewidth]{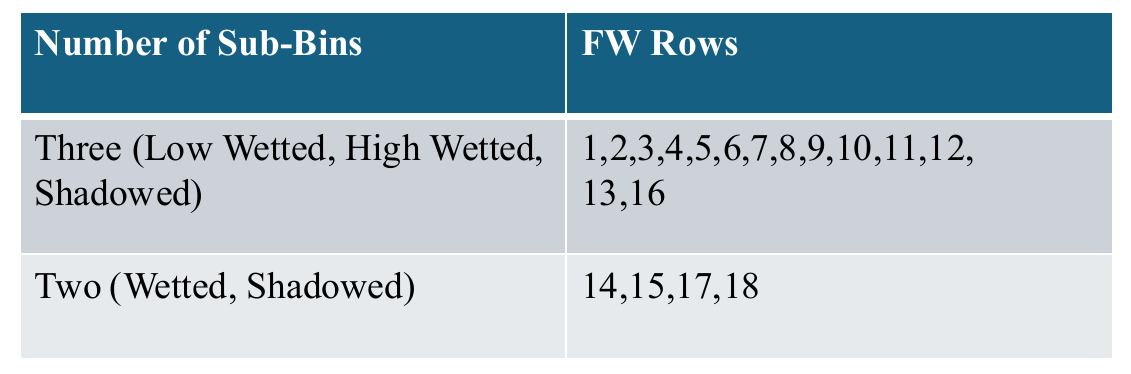}
    \caption{\gls{fw} sub-bins full breakdown.}
    \label{fig:fw-sub-bins}
\end{figure}

\begin{figure}[h!]
    \centering
    \includegraphics[width=\linewidth]{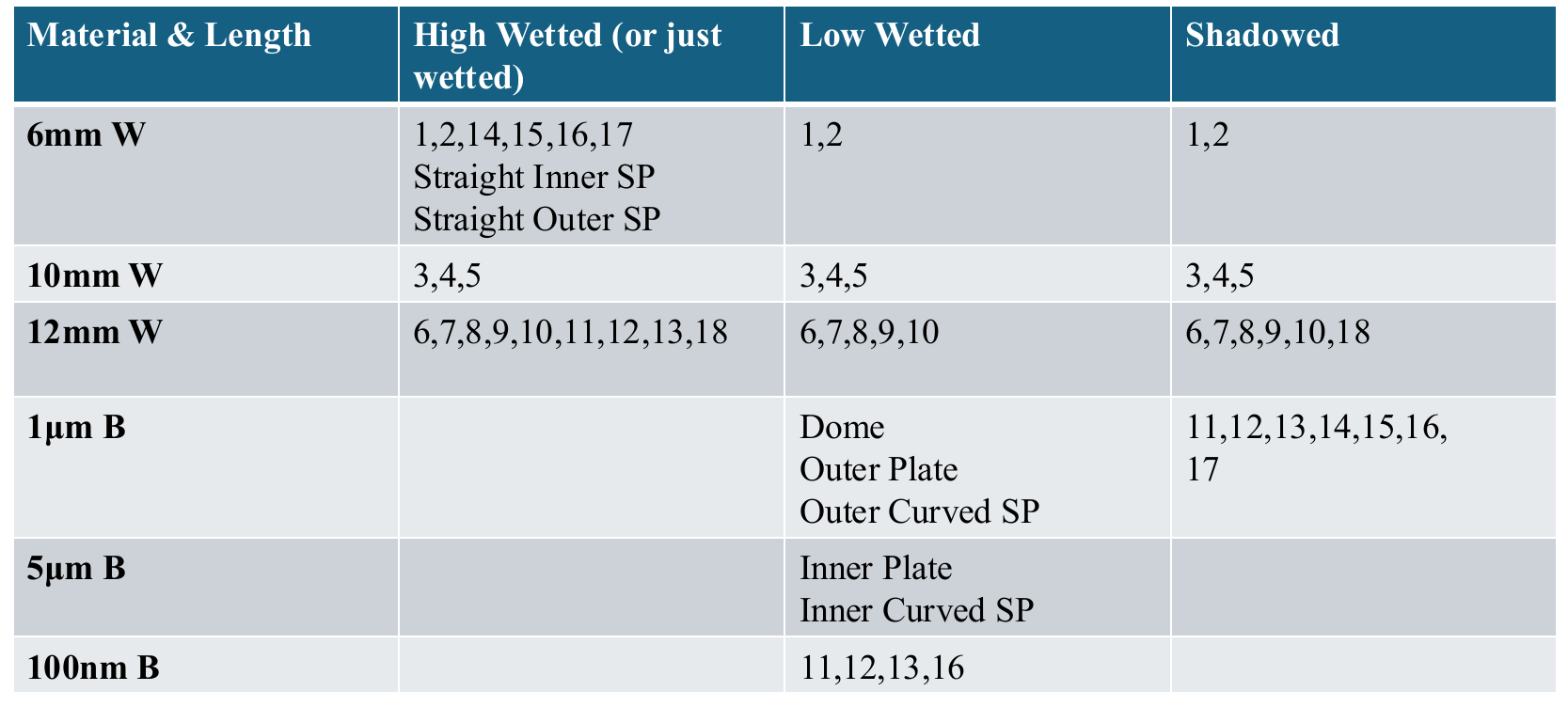}
    \caption{Main chamber and divertor bins by material and thickness.}
    \label{fig:bins-by-material}
\end{figure}

Figures \ref{fig:fw-sub-bins} and \ref{fig:bins-by-material} describe ITER bins by material, material thickness, and for the \gls{fw}, sub-bin count.

\section{Boron Model}
\label{app:Boron}
\begin{figure}[h!]
    \centering
    \includegraphics[width=0.5\textwidth]{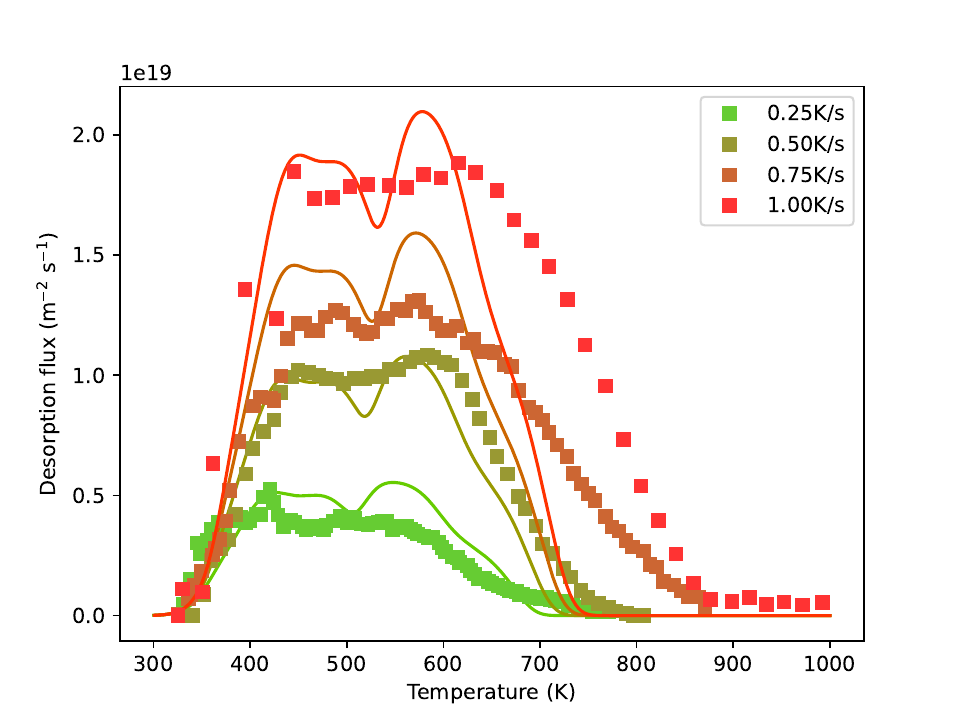}
    \caption{Comparison between experimental~\cite{OYA2004} (square markers) and simulated (solid line) TDS spectra for a thin boron film (47 nm) exposed to 1 keV D$_2^+$ ions at 304 followed by TDS at different heating ramps between 0.25~K s$^{-1}$ and 1~K s$^{-1}$.}
    \label{fig:TDS_Boron}
\end{figure}

\indent The model used to simulate D/T retention in the boron bins is validated by comparison with the results presented here.
In a study by Oya~\textit{et al}, a boron thin film (47 nm) is exposed to 1 keV D$_2^{+}$ ions at 312~K with a fluence up to 5.4$\times10^{21}$ D m$^{-2}$.
The deuterium desorption behavior is obtained by doing thermal desorption spectrometry measurements with a heating rate between 0.25 K s$^{-1}$ and 1 K s$^{-1}$.
These steps are simulated with the model reported in Tables~\ref{tab:boron} and \ref{tab:borontraps} using FESTIM.
The comparison between the experimental and simulated TDS spectra with a heating ramp between 0.25~K s$^{-1}$ and 1 K s$^{-1}$ is reported in figure~\ref{fig:TDS_Boron}, demonstrating a satisfactory simulation of the experimental data with the lowest relative error reaching down to 17\% (for 0.5 K s$^{-1}$).
A sensitivity analysis was carried out on the activation energy for diffusion by varying it between 0.3 eV and 0.5 eV. 
It shows no impact on the simulated TDS and the 0.3 eV value was chosen for the rest of the simulations (see Table~\ref{tab:boron}).
The model was first fitted to the low heating ramps as the integral of the high heating ramps. 1 K s$^{-1}$ is especially inconsistent with the incident fluence: it would lead to a retention rate higher than 100\% which is not physical.
However, this affects the high temperature part of the spectra with a missing high temperature contribution in the simulation with 0.75~K s$^{-1}$ and 1~K s$^{-1}$.

\bibliographystyle{unsrt}
\bibliography{references.bib}

\end{document}

%% file: glossary.tex
\newacronym[description=]{gdc}{GDC}{Glow Discharge Conditioning}

\newacronym[description=]{dd}{DD}{Deuterium-Deuterium}

\newacronym[description=]{dt}{DT}{Deuterium-Tritium}

\newacronym[description={one to four days of maintenance between full power operation where the torodial magnetic field can be stopped}]{stm}{STM}{Short Term Maintenance}

\newacronym[description=a weeks-long maintenance period following a complete operational campaign]{ltm}{LTM}{Long Term Maintenance}

\newacronym[description=]{fw}{FW}{First Wall}

\newacronym[description=]{icwc}{ICWC}{Ion Cyclotron Wall Conditioning}

\newacronym[description=]{risp}{RISP}{Raised Inner Strike Point}

\newglossaryentry{pulse}
{
    name=Pulse,
    description={An operational plasma defined by ramp up, ramp down, flat top (or steady state), and waiting times}
}

\newglossaryentry{day}
{
    name=Day,
    description={A full day of operation. One day can include multiple pulses and multiple kinds of pulses}
}

\newglossaryentry{cycle}
{
    name=Cycle,
    description={A fourteen-day sequence of operation. Cycles have full power operation days and maintenance days}
}

\newglossaryentry{campaign}
{
    name=Campaign,
    description={A months-long operational sequence, typically made of 8-10 cycles followed by a long-term maintenance period}
}

\newacronym[description=]{pfc}{PFC}{Plasma Facing Component}